\documentclass[11pt]{article}
\usepackage{moriond,epsfig}

\def\be{\begin{equation}}
\def\ee{\end{equation}}
\def\bea{\begin{eqnarray}}
\def\eea{\end{eqnarray}}

\begin{document}
\vspace*{4cm}
\title{LENSING SURVEY OF THE MOST X-RAY LUMINOUS GALAXY CLUSTERS}

\author{ W.KAUSCH (1), S. SCHINDLER (1), T. KRONBERGER (1), J. WAMBSGANSS (2), \\A. SCHWOPE (3), T. ERBEN (4) }

\address{(1) Institut f\"ur Astrophysik, Universit\"at Innsbruck,\\
Technikerstr. 25, A-6020 Innsbruck, Austria\\
(2) Institut f\"ur Physik, Universit\"at Potsdam, \\
Am Neuen Palais 10, D-14469 Potsdam, Germany\\
(3) Astrophysikalisches Institut Potsdam, \\
An der Sternwarte 16, D-14482 Potsdam, Germany\\
(4) Institut f\"ur Astrophysik und Extraterrestrische Forschung, Universit\"at Bonn,\\
Auf dem H\"ugel 71, D-53121 Bonn, Germany}

\maketitle\abstracts{
We present the first three galaxy clusters of a larger sample of the most X-ray luminous galaxy clusters
selected from the ROSAT Bright Survey. This project, which is a systematic search for strong lensing, aims at
arc statistics, mass determinations and studies of distant lensed galaxies.\\
The three galaxy clusters presented here have been observed with the Wide Field Imager at the ESO2.2m in the
R- and V-band. The images show lensing features like distinct distorted galaxies
and arcs. Mass distributions of the lensing galaxy clusters and photometric properties of some arc candidates
are presented. In addition we report the discovery of three giant arcs.}

\section{The Project}\label{sec:project}
The project is based on a well defined sample of 21 intermediate redshift ($0.10 < z < 0.52$) galaxy clusters.
These clusters are the most X-ray luminous systems taken from the ROSAT Bright Survey (Schwope et al.\cite{sc}).
As the X-ray luminosity of clusters is well correlated with their mass (Reiprich et al.\cite{reip}, Schindler,\cite{schin})
this sample is assumed to consist of very massive systems and hence the probability for them to act as gravitational
lenses is very high.\\
The goal of the project is a search for gravitational lensed objects, mainly arcs. These objects are highly
magnified images of very distant galaxies and hence they are useful probes for galaxy formation and their evolution.
In addition they can be used for investigations on the mass distribution and mass estimates of their lensing
clusters (see Bartelmann\&Schneider,\cite{bart2} for a review on gravitational lensing). Finally, the frequency of
giant arcs is a powerful tool to constrain and distinguish cosmological
models (Bartelmann et al.\cite{bart}, Wambsganss et al. \cite{wam}).\\
The observations were done with the SUperb Seeing Imager2 (SUSI2@ESONTT) and the Wide Field Imager (WFI@ESO2.2m)
in La Silla between August 2001 and April 2004. All clusters are observed in the R- and V-band, with the exposure
time in R usually being about twice that in V.\\
The data reduction is perfomed with the GaBoDS pipeline (Schirmer et al.\cite{sch}, Erben et al.\cite{er}).
This pipeline provides the data reduction in several steps: basic reduction (bias correction, overscan correction,
normal flatfielding), superflatting/defringing, astrometric and photometric calibration and, finally,
a coaddition is performed.
\section{WFI observations}\label{sec:WFI}
In this proceedings contribution we present the first three
clusters RBS325, RBS653 and RBS864. All observations were done at
good seeing conditions ($\leq 1"$) with the Wide Field Imager
during April 2002. In total there are 30 R-band images (ESO filter
BB\#Rc/162\_ESO844) with an exposure time $t_{exp}\sim$4.46h of
the coadded image, and 15 V-band images (BB\#V/89\_ESO843,
$t_{exp}\sim$2.23h) per cluster, except RBS653 where we have
obtained only 15 R-band images due to bad weather conditions
($t_{exp}\sim$2.23h in both R and V). The limiting magnitudes
above a $3\sigma$ detection threshold are given in
Tab.\ref{tab:properties}.
\section{Galaxy/Cluster Detection}\label{sec:galaxydetection}
\begin{figure}
\center \psfig{figure=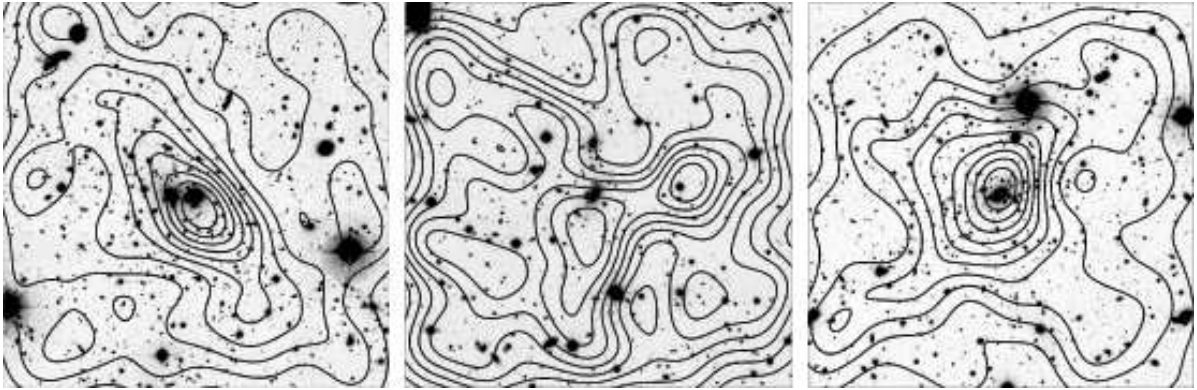,height=2.02in}
\caption{R-band images of RBS325 (left), RBS653 (middle) and
RBS864 (right). The Field of View is $6' \times 6'$ in all images
($\sim 2$Mpc $\times 2$ Mpc), north is up, east to the left. The
contours show the galaxy number densities of the Red Sequence (see
text for more details). \label{fig:galaxynumberplots}}
\end{figure}

The galaxy detection was performed with SExtractor 2.3.2 and a
3$\sigma$ detection limit. From the extracted catalogue we
selected all objects with the following properties: (a) objects
within a Field of View (FOV) of $1500\times 1500$ pixels ($\sim 6'
\times 6'$) centered on the cluster position given in the ROSAT
Bright Survey, (b) FWHM$_{object}>$ FWHM$_{stars}$, derived from
R$_{mag}$ vs. FWHM plots, and (c) CLASS\_STAR parameter of
SExtractor $\le$ 0.85, so we only select objects SExtractor
assumed to be galaxies.
\begin{table}[t]
\begin{center}
\begin{tabular}{|l|c|c|c|}
\hline
\bf property & \bf RBS325 & \bf RBS653 & \bf RBS864 \\
\hline
z & 0.282 & 0.286 & 0.2906 \\
$L_x$ [erg/sec] & 44.8 & 44.9 & 45.3 \\
FWHM$_{stars}$ (R\#813) [arcsec] & 0.95 & 1.30  & 1.15 \\
FWHM$_{stars}$ (V\#812) [arcsec] & 1.1 & 1.05 & 1.25 \\
TOTAL \# of detected galaxies & 1567 & 1285 & 1138 \\
limiting R-magnitude ($> 3.0\sigma$)& 25 & 25 & 25\\
limiting V-magnitude ($> 3.0\sigma$)& 25 & 24.5 & 25\\
\# of USED galaxies (see sec. \ref{sec:galaxydetection}) & 1083 & 1035 & 873  \\ \hline
\end{tabular}
\end{center}
\caption{summary of the cluster properties, $L_x=\log ($X-ray flux$)$.\label{tab:properties}}
\end{table}\newline
Tab. \ref{tab:properties} shows the properties of the three galaxy clusters and the constraints on the selected objects.
The resulting catalogues were used to extract the members of the foreground lensing cluster (the Red Sequence).
The galaxy number density plots (Fig. \ref{fig:galaxynumberplots}) were obtained by creating blank images of
the same size as the chosen Field of View ($\sim 6'\times6'$, $1500\times 1500$ pixels) and allocating pixel
value "1" to all positions referring to Red Sequence galaxies. This image is then strongly smoothed with
a $\sigma=300$ pixel Gaussian.
\section{Preliminary Results}\label{sec:results}
\subsection{Galaxy Number Density Plots}\label{subsec:galnumbdensplots}
Fig. \ref{fig:galaxynumberplots} presents the galaxy clusters RBS325, RBS653, and RBS864, the contours
represent the galaxy number densities of the corresponding Red Sequence.\\
RBS325 looks like a cluster which is at the start of a merging process: there is a significant alignment towards the
north-east direction in the number density plot, which could indicate the direction of a starting infall.
In contrast to that RBS653 shows no main direction in the galaxy number density plot. However, lots of substructures
can be seen, which indicates an advanced ongoing merger.
The galaxy number density plot of RBS864 presents only marginal substructures. This has been expected as a massive
cooling flow is known for this cluster (Edge et al.\cite{edge}).
\begin{figure}
\center \psfig{figure=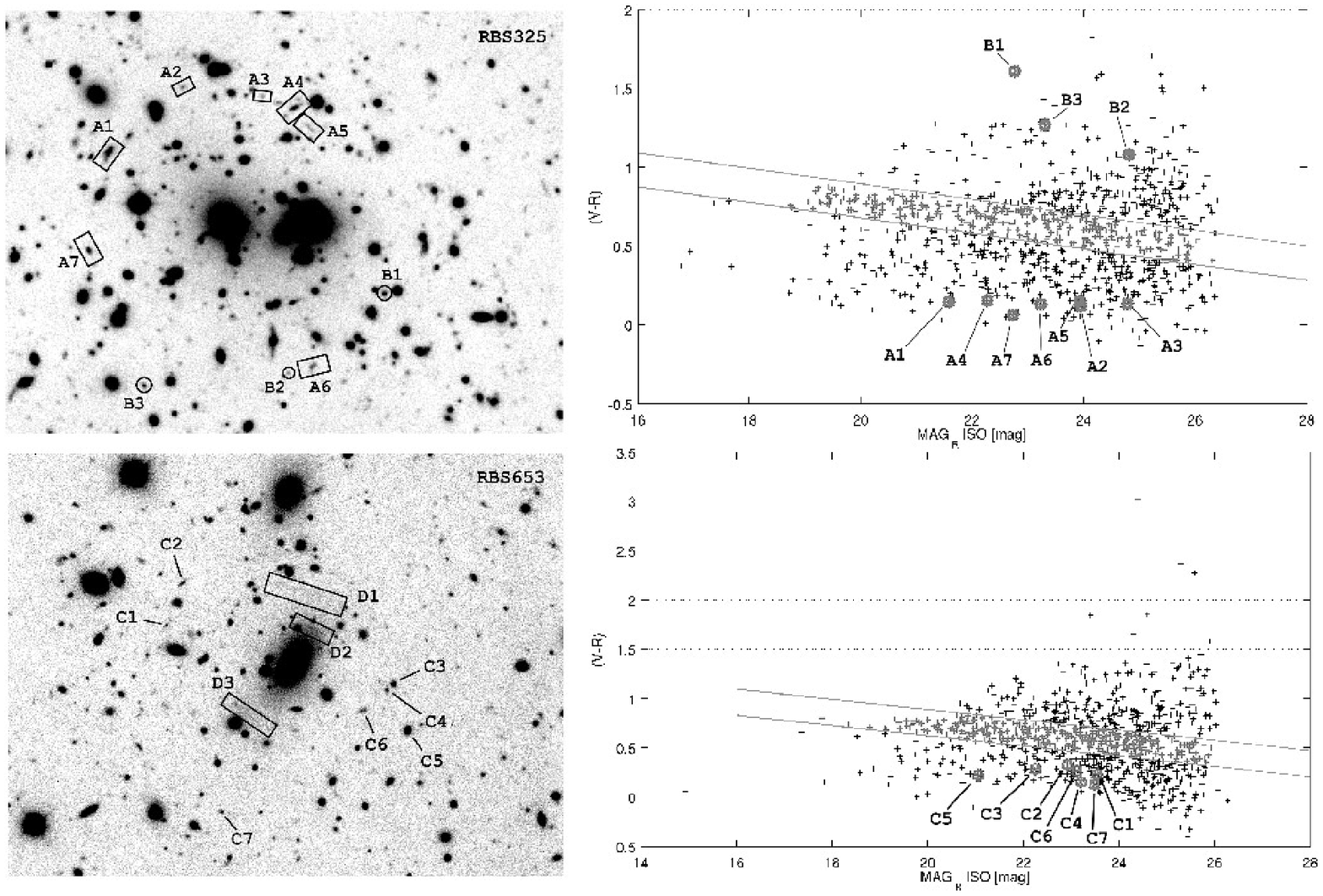,height=4.2in}
\caption{Left: center of the clusters RBS325 and RBS653 (FOV $\sim
2.5' \times 2') $. Right: color-magnitude plots for both clusters
(see text for more details)\label{fig:details}}
\end{figure}
\begin{table}
\begin{center}
\begin{tabular}{|l|c|c|c|c|c|c|c|c|c|c|}
\hline
\bf RBS325 & A1 & A2 & A3 & A4 & A5 & A6 & A7 & B1 & B2 & B3 \\[1mm]
\hline
(V-R) & 0.15 & 0.12 & 0.14 & 0.15 & 0.15 & 0.13 & 0.06 & 1.61 & 1.08 & 1.27\\
R mag& 21.6 & 24.0 & 24.8 & 22.3 & 24.0 & 23.2 & 22.7 & 22.8 & 24.8 & 23.3\\
\hline\hline
\bf RBS653 & C1 & C2 & C3 & C4 & C5 & C6 & C7 & - & - & -\\
\hline
(V-R) & 0.23 & 0.33 & 0.28 & 0.15 & 0.22 & 0.27 & 0.13 & - & - & -\\
R mag& 23.5 & 22.9 & 22.2 & 23.2 & 21.0 & 23.1 & 23.5 & - & - & - \\
\hline
\end{tabular}
\end{center}
\caption{Photometric properties of the objects marked in Fig. \ref{fig:details}\label{tab:results}}
\end{table}
\subsection{Photometry, Colours and Lensing Effects}\label{subsec:photometry}
As all our galaxy clusters are assumed to be very massive systems the probability for them to act as
gravitational lens is very high. Extrapolating from the EMSS numbers for galaxy clusters in different
X-ray luminosity ranges (Luppino et. al.\cite{lup}) we expect to find arc(lets) in 45\% of our clusters.
RBS864 shows no obvious strong lensing features, so we focus here on RBS325 and RBS653.\\
Fig. \ref{fig:details} left shows the centers of the clusters RBS325 and RBS653, the right part shows the
(V-R) vs. R-magnitude diagram with the Red Sequence galaxies between the two lines. The most interesting objects are
marked in both, their photometric properties are listed in Tab. \ref{tab:results}.\\
It can be seen that the objects marked with "A" in RBS325 nearly have the same color. In addition, A1, A5 and
A6 show a tangential alignment with respect to the cluster center. So the possibility that these are lensed
background objects is very high and will be checked by forthcoming observations. Objects B1, B2 and B3
are very red and possibly highly magnified very distant objects.\\
Also RBS653 shows several objects (marked with "C") with a different color index than the Red Sequence
(compare Tab. \ref{tab:results}). In addition
we report the discovery of several giant arcs, denoted by "D". The distances from the cluster center are roughly
$\sim$21" ($\sim$37 kpc, assuming $H_0=72 km s^{-1}Mpc^{-1}$), $\sim$10" ($\sim$18 kpc) and $\sim$20" ($\sim$35kpc),
their lengths are $\sim$16" ($\sim$28kpc), $\sim$7" ($\sim$12kpc)
and $\sim$9" ($\sim$16kpc) for D1, D2, and D3, respectively.\\[0.2cm]
{\bf Acknowledgement:} This work was funded by the Austrian Science Foundation (Fond zur F\"orderung der
wissenschaftlichen Forschung), Projectnumber P15868.

\end{document}